\documentclass{new_tlp}
\makeatletter
\gdef\@submitted{2 August 2016}
\gdef\@revised{2 August 2016}
\gdef\@accepted{?}
\makeatother

\usepackage{html}
\usepackage{ifthen}
\usepackage{calc}
\usepackage{mathptmx}
\usepackage{color}
\usepackage{listings}
\usepackage{multirow}
\usepackage[draft]{fixme}
\usepackage{graphicx}
\usepackage{fancyvrb}
\usepackage{color}
\usepackage{algpseudocode}
\usepackage{algorithm}
\usepackage{url}
\makeatletter
\g@addto@macro{\UrlBreaks}{\UrlOrds}
\makeatother
\catcode`\^^A=8
\catcode`\_=\active
\def_{\ifmmode\else\_\fi}

\definecolor{codeboxcolor}{rgb}{0.4,0.4,0.4}
\DefineVerbatimEnvironment%
  {code}{Verbatim}
  {frame=single,
   framerule=0.2pt,
   rulecolor=\color{codeboxcolor},
  }

\newcommand{\reffont}{\tt}

\newcommand{\predref}[2]{
        \mbox{\reffont #1/#2}}
\newcommand{\funcref}[2]{
        \mbox{{\reffont #1}(\textit{#2})}}

\newcommand{\const}[1]{
        \mbox{\reffont #1}}
\newcommand{\algoref}[1]{algorithm~\ref{algo:#1}}

\newcommand{\secref}[1]{section~\ref{sec:#1}}
\newcommand{\tabref}[1]{table~\ref{tab:#1}}

\newcommand{\figref}[1]{figure~\ref{fig:#1}}

\newcommand{\Tabref}[1]{Table~\ref{tab:#1}}

\title[Theory and Practice of Logic Programming]
      {Lock-free atom garbage collection \\ for multithreaded Prolog}

  \author[Jan Wielemaker and Keri Harris]
         {JAN WIELEMAKER\\
	  VU University Amsterdam, The Netherlands\\
	  CWI Amsterdam, The Netherlands\\
	  \email{J.Wielemaker@vu.nl}
	  \and KERI HARRIS\\
	  SecuritEase, New Zealand \\
	  \email{keri@gentoo.org}
	 }

\jdate{May 2016}
\pubyear{2016}
\pagerange{\pageref{firstpage}--\pageref{lastpage}}

\definecolor{lightgrey}{rgb}{0.95,0.95,0.95}

\newcommand{\jargon}[1]{\textit{#1}}

\makeatletter
\def\@nodescription{false}
\newcommand{\onlinebreak}{}

\newcommand{\defentry}[1]{\definition{#1}}
\newcommand{\definition}[1]{%
	\onlinebreak%
	\ifthenelse{\equal{\@nodescription}{true}}{%
	    \def\@nodescription{false}%
	    \makebox[-\leftmargin]{\mbox{}}\makebox[\linewidth+\leftmargin-1ex][l]{\bf #1}\\}{%
	    \item[{\makebox[\linewidth+\leftmargin-1ex][l]{#1}}]}}
\newcommand{\nodescription}{\def\@nodescription{true}}
\def\predatt#1{\hfill{\it\footnotesize[#1]}}
\def\predicate{\@ifnextchar[{\@attpredicate}{\@predicate}}
\def\qpredicate{\@ifnextchar[{\@attqpredicate}{\@qpredicate}}
\def\@predicate#1#2#3{%
	\ifthenelse{\equal{#2}{0}}{%
	    \defentry{#1}}{%
	    \defentry{#1({\it #3})}}%
	\index{#1/#2}\ignorespaces}
\def\@attpredicate[#1]#2#3#4{%
	\ifthenelse{\equal{#3}{0}}{%
	    \defentry{#2\predatt{#1}}}{%
	    \defentry{#2({\it #4})\predatt{#1}}}%
	\index{#2/#3}\ignorespaces}
\def\@qpredicate#1#2#3#4{%
	\ifthenelse{\equal{#3}{0}}{%
	    \defentry{#1:#2}}{%
	    \defentry{#1:#2({\it #4})}}%
	\index{#1/#2}\ignorespaces}
\def\@attqpredicate[#1]#2#3#4#5{%
	\ifthenelse{\equal{#4}{0}}{%
	    \defentry{#2:#3\predatt{#1}}}{%
	    \defentry{#2:#3({\it #5})\predatt{#1}}}%
	\index{#2/#3}\ignorespaces}
\def\directive{\@ifnextchar[{\@attdirective}{\@directive}}
\def\@directive#1#2#3{%
	\ifthenelse{\equal{#2}{0}}{%
	    \defentry{:- #1}}{%
	    \defentry{:- #1({\it #3})}}%
	\index{#1/#2}\ignorespaces}
\def\@attdirective[#1]#2#3#4{%
	\ifthenelse{\equal{#3}{0}}{%
	    \defentry{:- #2\predatt{#1}}}{%
	    \defentry{:- #2({\it #4})\predatt{#1}}}%
	\index{#2/#3}\ignorespaces}
\newcommand{\termitem}[2]{%
	\ifthenelse{\equal{}{#2}}{%
	    \definition{#1}}{%
	    \definition{#1({\it #2})}}\ignorespaces}
\makeatother

\begin{document}

\label{firstpage}

\maketitle

  \begin{abstract}
The runtime system of dynamic languages such as Prolog or Lisp and their
derivatives contain a \textit{symbol table}, in Prolog often called the
\textit{atom table}. A simple dynamically resizing hash-table used to be
an adequate way to implement this table. As Prolog becomes fashionable
for $24 \times 7$ server processes we need to deal with atom garbage
collection and concurrent access to the atom table. Classical lock-based
implementations to ensure consistency of the atom table scale poorly and
a stop-the-world approach to implement atom garbage collection quickly
becomes a bottle-neck, making Prolog unsuitable for soft real-time
applications. In this article we describe a novel implementation for the
atom table using lock-free techniques where the atom-table remains
accessible even during atom garbage collection.  Relying only on CAS
(Compare And Swap) and not on external libraries, the implementation
is straightforward and portable.

Under consideration for acceptance in TPLP.
  \end{abstract}



\section{Introduction}

An important, but for a long time simple, component of the
implementation of dynamic languages is the symbol or atom table. This
table maps the string representation of a symbol (atom) to a
\textit{handle}. Using handles instead of the original strings avoids
duplication, allows for fast equality testing (unification, clause
indexing) and causes each symbol to require the same space, which
simplifies the representation of data involving atoms. The atom table is
traditionally implemented using a dynamically resizing \jargon{open hash
table}. Such a table provides excellent performance and is easy to
implement.

In languages that use symbols to represent only constants from the
program, the above is completely adequate. Prolog programs, however,
tend to use atoms also for representing constant as well as dynamic
strings from data that is processed by the program. For example, NLP
(natural language processing) programs tend to represent words from the
text they process as atoms. As shown in, e.g.,
\citeN{Creutz:2003:USW:1075096.1075132}, the number of unique words does
not flatten out if more and more documents are being processed. This
means we need \jargon{atom garbage collection} (AGC) to dispose of words
from old documents if we want to realise programs that can process
unbounded input.

As virtually all modern hardware has multiple cores and the number of
cores is growing, several Prolog implementations have added support for
multiple threads (see \tabref{agcsupport}). In most systems a thread
runs a \jargon{goal} using its own stacks, while all threads share the
same \jargon{program}. Typically, atoms are shared between all threads
and thus access to the atom table needs to be synchronised between
threads.

A naive way to solve this problem is described in
\citeN{DBLP:conf/iclp/Wielemaker03}. It relies on a classical atom table
for which the consistency is guaranteed using \jargon{mutexes} (also
called \jargon{locks} or \jargon{critical sections}) that serialises
atom lookup operations. Note that an atom lookup may either return an
existing atom or create a new atom. In a typical application most atom
lookup operations return an existing atom. Still, also lookup of an
existing atom needs to be locked to deal with a table resize as well as
AGC. The first implementation of concurrent AGC in SWI-Prolog used a
\jargon{stop-the-world} approach: the thread that initiates AGC stops
all threads, marks all atoms reachable from each thread, removes
unmarked atoms from the hash table and finally resumes all threads. This
became problematic for two reasons. First, we discovered that reliably
`stopping the world' is troublesome in MS~Windows.\footnote{This is
claimed
(\url{http://www.codeproject.com/Articles/7238/QueueUserAPCEx-Version-Truly-Asynchronous-User-M})
to be possible using a device driver. Using a device driver was not an
option due to the required administrative privileges for installing a
device driver as well as security considerations that cause many
organizations not to accept software that require this. Another claim
found is to use GetThreadContext() after SuspendThread(). This proved
unreliable in our tests (around 2011) on Windows XP. Similar problems
are reported at e.g.,
\url{http://stackoverflow.com/questions/3444190/windows-suspendthread-doesnt-getthreadcontext-fails}.
Microsoft hints at this solution in a recent (2015) post at
\url{https://blogs.msdn.microsoft.com/oldnewthing/20150205-00/?p=44743}.}
Second, as programs relying on dozens of threads were developed, the
atom table lock became a bottleneck. Our requirements are:

\begin{itemize}
    \item AGC that allows other threads to proceed,
          including performing atom lookups.  This solves the above
	  mentioned portability problem and makes the system better
	  suitable for (soft) real-time applications.
    \item Scalable atom lookup, where the lookup time depends as
          little as possible on the number of threads performing
	  concurrent lookups.
\end{itemize}

The first step to tackle this was taken in 2013 after we discovered
the portability issue mentioned above. We replaced the stop-the-world
collector with an asynchronous marking algorithm while using the global
atom table lock to perform the collection phase safely. This implies
that other threads can proceed during the marking phase, but will block
when trying to lookup an atom. The design of this collector is the
subject of \secref{conservative-agc}.


With portable support for \jargon{atomic} memory operations, notably
compare-and-swap (CAS) now being available for all major platforms as
well as a wealth of described techniques for using these to build
lock-free data structures (e.g., \jargon{Read-copy-update} (RCU),
\jargon{Transactional Memory} (TM), \jargon{Hazard pointers}
\cite{10.1109/TPDS.2011.159,DBLP:journals/cacm/HarrisMJH08,1291819}) we
decided to replace many of the shared data structures in SWI-Prolog with
lock-free alternatives. A crucial and the most challenging data
structure was the atom table. The design of this lock-free, resizing
hash table and its synchronisation with AGC is the subject of
\secref{atomtable}.

This article is organised as follows. In \secref{related} we discuss the
state-of-the-art with regard to atom handling in Prolog and other
related work. In \secref{agc} and \secref{atomtable} we describe our
implementation of the atom garbage collector and the lock-free atom
table. In \secref{evaluation} we evaluate our implementation using a
couple of real applications as well as an artificial benchmark.

\section{Related work}
\label{sec:related}

Atom garbage collection is still not widespread in Prolog nor
related languages such as Erlang or Ruby. Ruby supports symbol
garbage collection as of version
2.2.\footnote{\url{https://www.infoq.com/news/2014/12/ruby-2.2.0-released}}
To our best knowledge, Erlang does not provide atom/symbol
garbage collection. \Tabref{agcsupport} summarises the support for
threads and AGC in popular Prolog systems. AGC for Erlang has been
proposed in \citeN{Lindgren:2005:AGC:1088361.1088369} based on the same
motivation as we have, atom are commonly used for the representation of
data that is being processed and long running processes will thus
collect too many atoms over time. The Erlang community deals with this
by avoiding atoms, using lists of characters instead or by restarting
nodes periodically. The Prolog community uses the same workarounds. Some
systems, e.g., SWI-Prolog, ECLiPSe and LPA Prolog support packed strings
to have a compact and natural representation for volatile text as well
as avoid the need for AGC.

\begin{table}[htb]
\begin{tabular}{lccp{0.6\linewidth}}
\textbf{System}	& \textbf{Threads} & \textbf{AGC} & \textbf{Notes} \\
\hline
SWI-Prolog         & Y      & Y     \\
SICStus 4	   & N      & Y    & \footnotesize
				     SICStus supports multiple independent
				     runtime systems in one process \\
YAP 6.3		   & Y      & Y    & \footnotesize
				     YAP supports AGC or threads, but not both \\
B-Prolog 8.1       & N      & N     \\
CxProlog 0.98.1    & N      & Y     \\
ECLiPSe 6.1        & N      & Y    & \footnotesize
				     ECLiPSe will soon support threads and AGC.  The
				     implementation is similar to
				     \citeN{Lindgren:2005:AGC:1088361.1088369} and
				     \citeN{Tarau:2011:IST:1993478.1993497} \\
GNU Prolog 1.4.4   & N      & N     \\
JIProlog 4.1.4.1   & N      & N     \\
Lean Prolog 5.4.4  & Y      & Y    & \footnotesize
				     See \secref{related} \\
Qu-Prolog 9.7      & Y      & N     \\
XSB 3.6.0          & Y      & N     \\
\end{tabular}
    \caption{Thread and AGC support for some popular Prolog systems}
    \label{tab:agcsupport}
\end{table}

We find two types of related work in the literature. First, there is a
quickly growing body of articles describing lock-free data structures.
We particularly refer to \citeN{DBLP:conf/usenix/TriplettMW11},
describing a lock-free hash tables based on kernel space RCU techniques.
This paper has an extensive section on related techniques for lock-free
hash tables. The hash table implementation described provides good
lookup performance during resize, a property lacking in our
implementation (see \secref{atomtable}). The downside is that it relies
heavily on the RCU \jargon{wait-for-readers} action, the implementation
of which is slow in user space and poorly portable. We considered using
\texttt{liburcu}\footnote{\url{http://liburcu.org}}, but discarded it
due to the lack of support for native MS-Windows as well the
lack of portability in general that follows from the detailed list of
supported CPUs and compilers.

Second, we looked at the work done in the area of multi-threaded
symbolic languages. In \citeN{Lindgren:2005:AGC:1088361.1088369}, Thomas
Lindgren describes the design of an atom garbage collector for Erlang.
The overall idea is to realise a \jargon{copying} collector where we
have two symbol tables. If AGC is started, a new table is initialised
with the permanent atoms. An Erlang process is moved to the current
(new) table when it is resumed. The process scans its memory areas and
moves each atom to the new table while updating the used atom-handle. If
all processes have been moved, the old table can be discarded.
\citeN{Tarau:2011:IST:1993478.1993497} describes the symbol
garbage collector for \textit{Lean Prolog}. This Java based minimalist
Prolog implementation uses symbols as a generalisation for atoms that
can also refer to Java objects such as large numbers.\footnote{Also
SWI-Prolog atoms are internally generalised to symbols that are also used as
safe references to \jargon{foreign} objects such as streams, clause
references, etc.} The implementation follows the same copying approach
as the Erlang proposal described above.

The copying approach has two advantages. First, the size of the atom
table is actually reduced and second, migrating the atoms is done by the
target thread itself, which avoids the need for asynchronous scanning as
described in \secref{conservative-agc} and naturally distributes the
workload over the running threads. In our view, there are also
disadvantages. First, after starting a new symbol table, it needs to be
populated with all permanent atoms, e.g., those appearing in the
static part of the program. Second, all threads will concurrently
populate the new table with a potentially large number of atoms. Third,
SWI-Prolog threads are particularly designed to be embedded into C code
and call arbitrary C code. This may cause long (even infinite) delays
before all threads have migrated to the new table and we can discard the
old one. This issue is raised by Lindgren but not resolved. Tarau does
not mention this. Third, SWI-Prolog is actively used for processing
\jargon{linked data} (RDF, \citeNP{rdfconcepts}). Applications like this
use many atoms (we used up to 50~million atoms). Copying these atoms is
expensive and uses a large amount of memory.

The collector described in this article allows threads to proceed their
work during AGC, which includes running Prolog code, creating atoms as
well as being blocked in system calls or expensive computations done in
external languages. It reclaims the actual strings of unreachable atoms.
The atom itself is not reclaimed, but reused when new atoms need to be
created. It is not hard to imagine workloads where the copying approach
is preferable to merely reusing atoms, neither the other way around. It
is hard to make a fair judgement for real world usage.

\section{A conservative atom garbage collector}
\label{sec:agc}

This section describes AGC as it is currently implemented in SWI-Prolog.
We briefly describe the simple single threaded and stop-the-world
algorithms before introducing our current asynchronous marking
algorithm. For AGC purposes we distinguish two types of references to
atoms.

\begin{enumerate}
    \item \jargon{Volatile} references come from highly dynamic
    memory areas.  Currently these are the Prolog stacks (global
    and environment), the buffer area used by findall/3 and terms
    in \jargon{message queues} (streams of terms used to exchange
    messages between threads). Atoms referenced from these areas are
    identified by scanning these areas during the \jargon{mark} phase of
    AGC.

    \item \jargon{Explicit} references come from mostly
    static data structures such as clauses, records, Prolog flags
    and code using the C interface.  The number of such references
    is stored with the atom and maintained using
    \funcref{PL_register_atom}{} and \funcref{PL_unregister_atom}{}.
\end{enumerate}

The atom lookup functions (\funcref{PL_new_atom}{} and variations)
increment the reference count of the returned atom to avoid it being
collected immediately after creation. Functions such as
\funcref{PL_put_atom_chars}{}, which bind a Prolog term to an atom
created from a string call \funcref{PL_new_atom}{}, bind the term to the
atom and decrements the reference count of the atom. Thus, calling
\funcref{PL_put_atom_chars}{} with a new unique string creates an atom
that is referenced from a term and has its \const{references} field set
to zero. The highest bit of the \const{reference} (\const{marked})
field is used for marking that there is a reference from a volatile
memory area. Now, AGC performs the steps outlined in \algoref{simpleagc}

\begin{algorithm}[htb]
\begin{algorithmic}[1]
\footnotesize
\Function{agc}{}
  \ForAll{$volatile\_area$}
    \State $\Call{mark_atoms_in}{volatile\_area}$
  \EndFor
  \State
  \ForAll{$a$ \textbf{in} $atoms$}
    \If{$a.references = 0$}\Comment{no mark, no explicit rereferences}
      \State $\Call{reclaim_atom}{a}$
    \Else
      \State $\Call{clear_mark}{a}$
    \EndIf
  \EndFor
\EndFunction
\end{algorithmic}
    \caption{Simple AGC}
    \label{algo:simpleagc}
\end{algorithm}

The above works for single-threaded execution. If we add threads to the
picture, we need to take care of volatile references from other threads.
We describe two approaches for this. First we provide a brief
description of our old stop-the-world collector
\cite{DBLP:conf/iclp/Wielemaker03}, followed by a description of our
current conservative collector.

\subsection{Stop the world AGC}
\label{sec:stop-the-world-agc}

Stop-the-world agc is similar to the single-threaded of
\algoref{simpleagc}. It merely has to mark the volatile areas of all
threads. To do so, it stops each thread and marks all atoms reachable
from the volatile areas of the stopped thread. Next, it collects the
unreachable atoms and finally it resumes the stopped threads. Note that
we cannot resume the threads immediately after marking because they may
add new atoms to their volatile areas.

On Unix systems, each thread is signalled. The signal handler marks the
reachable atoms and then suspends using \funcref{sigwait}{}. On Windows,
threads are stopped using \funcref{SuspendThread}{}, after which the AGC
thread marks the atoms of the suspended thread. Later we discovered that
\funcref{SuspendThread}{} returns immediately and only prevents the
target thread from resuming after its current time slice finishes. After
many workaround attempts we concluded there is no reliable way to
suspend a thread and wait for it to be really suspended.

Although only one thread is accessing the thread's stacks during
marking, the marking happens asynchronously as to avoid AGC (and thus
all threads) having to wait until all threads reach a safe check
point. This requires careful ordering of modifications to the stacks.

\subsection{Conservative AGC}
\label{sec:conservative-agc}

Blocking threads during the entire AGC process, the requirement to scan
the stacks asynchronously and the portability issue around suspending
threads were the major reasons to seek for another solution for marking
the volatile areas. The inspiration came from the
\textit{Boehm-Demers-Weiser garbage collector} (BDWGC,
\citeNP{Boehm:1993:SEC:155090.155109}) which performs garbage
collection on C data structures by scanning all memory for values that
\emph{can} be interpreted as a pointer to a location inside an allocated
block of memory. Otherwise, BDWGC is a stop-the-world collector.

Instead of obtaining a root pointer to the current environment frame and
choice point and examining all reachable frames and atoms from there, we
simply scan the environment and global stack and mark anything that
looks `atom-like', but can of course be an accidental bit pattern
appearing in, e.g., a floating point number or string. This is the
\jargon{conservative} aspect: the marker might mark atoms that are in
fact not referenced and thus the collector might not collect all atoms.
The probability for false marks is reduced by using a tag on the lower
bits of an atom handle that excludes clashes with aligned pointers.

As we do not want to stop threads, we need to deal with the fact that
the target thread is running and changing the stacks as we mark them.
Because we merely examine the cells in the volatile areas, trying to
interpret the bit pattern as an atom, our marker is safe as long as the
location of the volatile areas of a thread remains unchanged while the
area is scanned and atoms in the area are not somehow made invisible or
moved. The location of volatile areas is changed during a stack shift,
while atoms can be invisible and be moved during a thread-local
stack-GC. Therefore these two operations needs to be synchronized
between the target thread and the marking thread using locks.

\begin{figure}[htb]
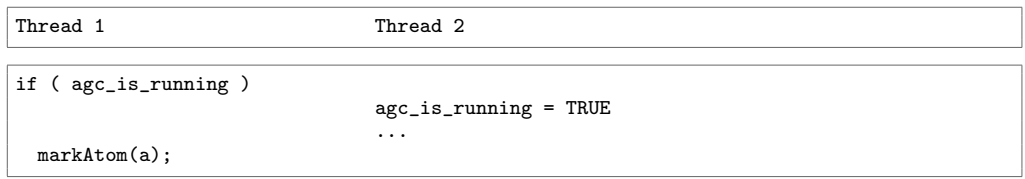

\footnotesize
\begin{code}
Thread 1			Thread 2
\end{code}
\begin{code}
if ( agc_is_running )
				agc_is_running = TRUE
			        ...
  markAtom(a);
\end{code}
    \caption{Race when conditional marking is used}
    \label{fig:atom-race}
\end{figure}

The running thread may create volatile references to new atoms as it
proceeds, e.g., by pushing an atom to a stack. Without precaution, this
atoms may not be marked. Therefore, we introduce a global variable to
indicate that AGC is in progress. When AGC is in progress, operations
that add an atom to one of the volatile areas also mark the atom.
Similarly, \funcref{PL_unregister_atom}{} must mark an atom if the last
reference is lost while AGC is in progress. A simple conditional mark is
insufficient due to the race condition illustrated in
\figref{atom-race}. This is solved by placing such atoms in a designated
field in the thread structure that is marked by AGC as illustrated in
\algoref{atom-race-fix}. In this figure, \const{LD} represents the
thread structure. Now, the atom is marked either by AGC thread scanning
\const{LD->atoms.unregistering} or by the calling thread.

\begin{algorithm}[htb]
\begin{algorithmic}[1]
\footnotesize
\Function{cond_mark_atom}{$a$}
  \State $LD.atoms.unregistering \gets a$
  \If{$agc\_is\_running = true$}
    \State \Call{mark_atom}{a};
  \EndIf
\EndFunction
\end{algorithmic}
    \caption{Safe conditional marking}
    \label{algo:atom-race-fix}
\end{algorithm}

With these changes the AGC implementation becomes as illustrated in
\algoref{cagc}. This implementation has the following properties:

\begin{itemize}
    \item Threads continue during AGC marking.  Currently, AGC is
    performed by the initiating threads.  Future versions may pass
    this to a dedicated thread and may use multiple threads for
    the marking.
    \item Threads suspend on stack shifts or garbage collection while
    their volatile areas are being scanned.  This is acceptable because
    the additional marking delay is proportional to the delay involved
    with GC or stacks shifts.  More fine grained locking, e.g., by
    volatile area, is possible.
    \item Threads creation and destruction suspends during the
    marking phase of AGC.  This also allows for more fine grained
    locking.
    \item Atom lookup suspends during the collect phase.
    \item No non-portable constructs such as suspending other
    threads are needed.  The algorithm can be fully implemented
    using \textsc{posix} thread primitives, although our implementation
    uses atomic operations for, e.g., updating the atom reference count.
\end{itemize}

\begin{algorithm}[htb]
\begin{algorithmic}[1]
\footnotesize
\Function{agc}{}
  \State \Call{lock}{agc}
  \If{$agc\_is\_running = true$}
    \State $\Call{unlock}{agc}$
    \Return
  \EndIf
  \State $agc\_is\_running \gets true$
  \ForAll{$t$ \textbf{in} $threads$}
    \State $\Call{mark_volatile}{t}$
  \EndFor
  \State \Call{lock}{atom_table}
  \ForAll{$a$ \textbf{in} $atoms$}
    \If{$a.references = 0$}
      \State $\Call{reclaim_atom}{a}$
    \Else
      \State $\Call{unmark}{a}$
    \EndIf
  \EndFor
  \State \Call{unlock}{atom_table}
  \State $agc\_is\_running \gets false$
  \State $\Call{unlock}{agc}$
\EndFunction
\end{algorithmic}
    \caption{Conservative AGC control}
    \label{algo:cagc}
\end{algorithm}

\section{A lock free atom table}
\label{sec:atomtable}

Although the in \secref{conservative-agc} described atom garbage
collector improves portability and reduces the time in which no thread
can make progress, it does not avoid contention on the atom table lock
and it still causes thread doing atom lookup to block during the collect
phase. We describe our implementation that solves these problems in this
section. We make the following assumptions and have the following
requirements:

\begin{itemize}
    \item A particular application requires up to a certain number
    of \emph{live} atoms. This number is not known in advance and
    therefore the atom table needs to be resized until the required
    size is reached.
    \item Although it is necessary to remove no-longer-used atoms
    from the table, there is not much need to \emph{reduce} the
    number of buckets in the atom table.  This implies that after a
    startup period, atom table resize operations no longer take
    place and thus rather poor atom handling performance during
    the resize operations is acceptable.
    \item However, atom garbage collection may remain a frequent
    activity and thus atom lookup must perform well during AGC.
    \item The implementation must be portable to major operating
    systems and CPUs.  In particular, we wish to limit the required
    synchronisation primitives to the POSIX mutexes (critical
    sections on Windows) and the atomic \jargon{Compare And
    Swap} (CAS) operation for pointers and integers up to the
    size of a pointer.  Modern C~compilers generally provide CAS
    through documented primitives such as GCC's
    \funcref{__sync_bool_compare_and_swap}{ptr,old,new} rather
    than relying on embedded assembly code.
\end{itemize}

\subsection{Data structures}
\label{atom-data}

\begin{figure}[htb]
\includegraphics[width=0.7\linewidth]{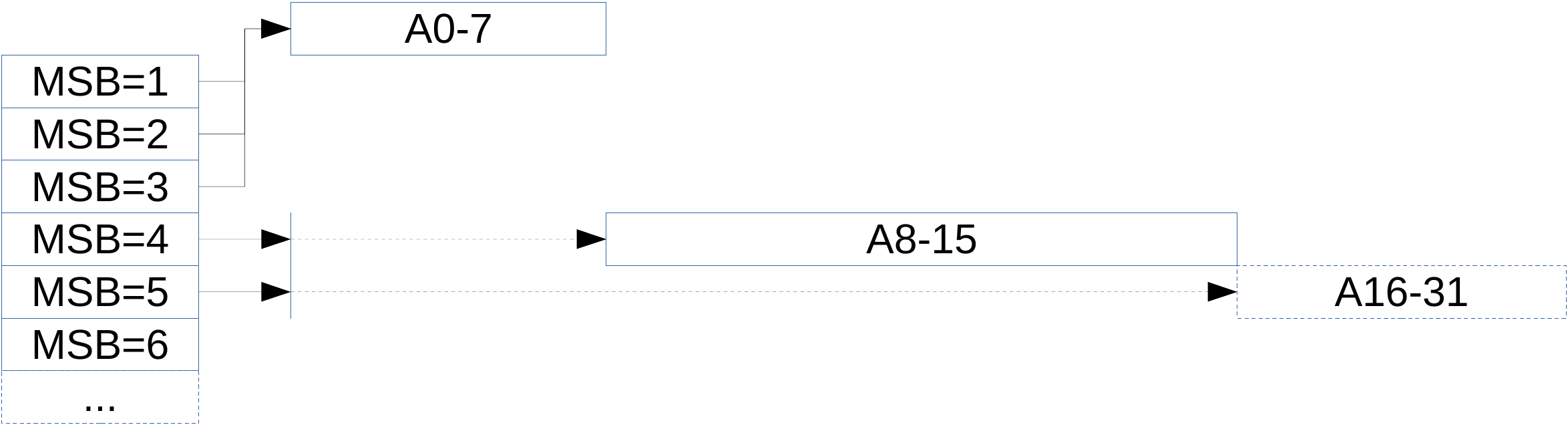}
\caption{Dynamic array data structure}
\label{fig:dynarray}
\end{figure}

The atom-table consists of a dynamic array of atom structures. The
dynamic array is implemented using an array of base pointers to chunks
that double in size as illustrated in \figref{dynarray}. Typically, the
array is statically initialized to a specified size (8 in the figure).
Using this representation, the atom at index $I$ ($I > 0$) can be
requested using $atomArray[\Call{MSB}{I}][I]$.\footnote{Many CPUs
provide hardware support to compute the \jargon{Most Significant Bit}.
For example, GCC provides access to this using
\funcref{__builtin_clzl}{}.} The dynamic array can be extended by
allocating a new block and adding it to the MSB index.

The atom structures in the dynamic array have the fields described
below. In the actual implementation they have more fields, but these are
not relevant for our description of the atom garbage collector.

\begin{description}
    \item [next] Pointer to the next atom in the open hash table.
    \item [name] Pointer to the represented text.
    \item [references] The atom reference
    count.  The top three bits are named \const{reserved},
    \const{valid} and \const{marked}.  The \const{marked} bit
    is used for marking references from volatile areas, the
    \const{reserved} bit is used to indicate that the atom is
    not available for creating a new atom and the
    \const{valid} bit indicates the atom is fully alive.
    \item [next_invalid] Pointer to next invalidated (but not yet
    reclaimed) atom.  The use of this field is clarified in
    \algoref{invalidateatom}.
\end{description}

\begin{figure}[htb]
\includegraphics[width=0.7\linewidth]{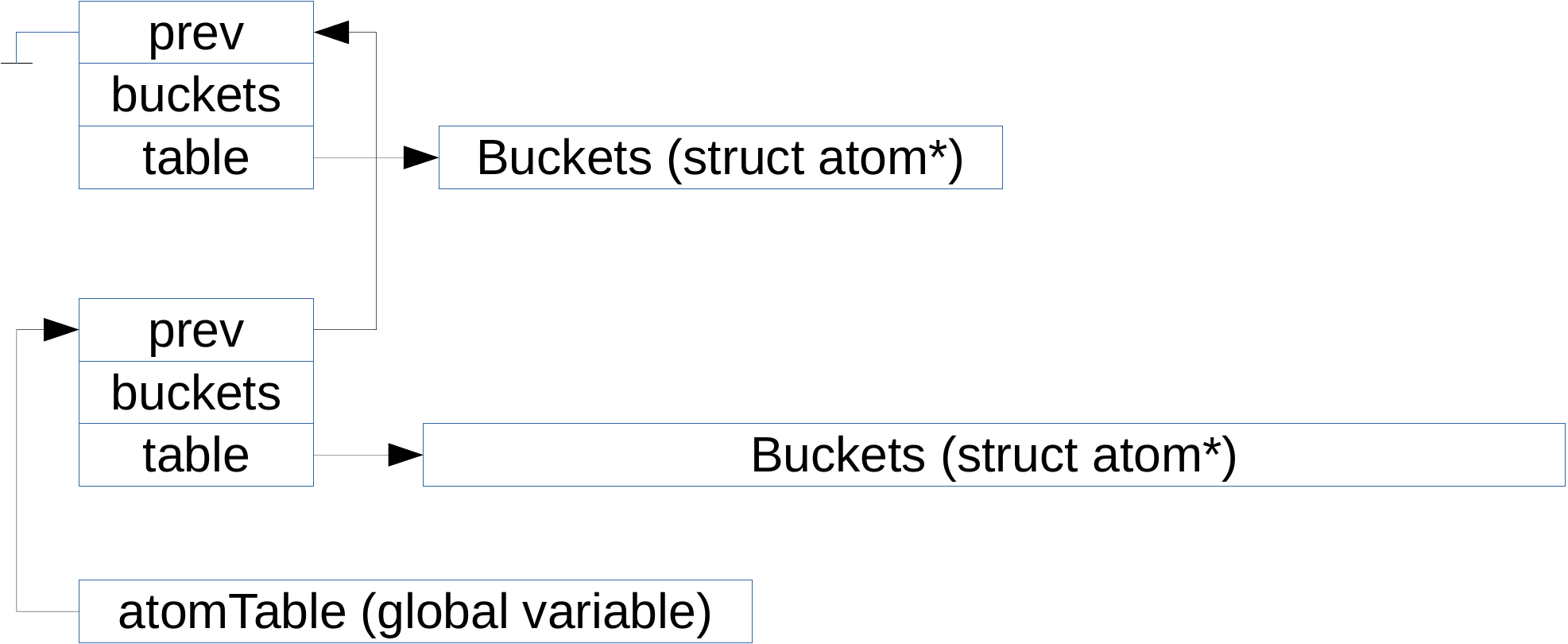}
\caption{The atom_table structure}
\label{fig:atomtable}
\end{figure}

The \const{atom_table} is a classical open hash table. Following the RCU
approach, the atom table is represented using a structure that is
atomically replaced by a new (resized) version by making the global
\const{atomTable} pointer point at the new version. Old versions remain
reachable through the \const{prev} pointer until they can safely be
reclaimed. Reclaiming old structures is described in
\secref{algorithm}.  See \figref{atomtable}.

\subsection{Algorithm}
\label{sec:algorithm}

This section describes the algorithm to manage the atom hash table as
well as reclaiming atoms from AGC. Note that that AGC mark phase
described in \secref{conservative-agc} is not affected. The algorithm is
provided as pseudo code. The description below summarises the algorithm
while pointing at the relevant fragments of the pseudo code
implementation.

\begin{algorithm}[htb]
\begin{algorithmic}[1]
\footnotesize
\State \textbf{global var} $atomTable$
\State \textbf{thread var} $LD$
\State

\Function{lookup_atom}{string}
  \State \textbf{var} $table, buckets, head$

  \Loop
    \State $LD.atomTable = atomTable$
    \State $table \gets LD.atomTable.table$
    \State $buckets \gets LD.atomTable.buckets$
    \State $key \gets \Call{hash}{string} \& (buckets-1)$
    \State $head \gets table[key]$
    \State $LD.atomBucket = \&table[key]$

    \State
    \For{$a \gets head; a; a \gets a.next$}
      \State $references \gets a.references$
      \If{$\Call{is_valid}{references} \bigwedge string = a.name$}
        \If{$\Call{bump_ref}{a, references}$}
	  \State $LD.atomBucket = LD.atomTable = NULL$
	  \State \Return $a$
	\EndIf
      \EndIf
    \EndFor

    \State
    \If{table too full}
      \State \Call{resize_atom_table}{}
    \EndIf

    \If{$table$ or $bucket$ not current}
      \State \textbf{continue} loop
    \EndIf

    \State
    \State $a \gets \Call{reserve_atom}{string}$
    \State $a.next \gets table[key]$
    \If{$\Call{cas}{\&table[key], head, a} \bigwedge
	table$ is current}
      \State $a.references \gets 1|VALID|RESERVED$
      \State $LD.atomBucket = LD.atomTable = NULL$
      \State \Return $a$
    \Else
      \State $a.references \gets 0$
    \EndIf
  \EndLoop
\EndFunction
\end{algorithmic}

    \caption{Atom lookup}
    \label{algo:atomlookup}
\end{algorithm}

\begin{enumerate}
    \item Fetch the current global atom-table and do a classical open
    hash table lookup.  If the atom is found and marked \const{valid},
    increment \const{references} using CAS while validating that the
    atom remains marked as \const{valid} and return the atom.
    See \algoref{atomlookup}, lines 7 to 21 and \algoref{bumpref}.

    \item If the table is too small (\algoref{atomlookup}, lines 24 to 26),
    resize the table (\algoref{resizeatomtable}). While the resize
    is in progress, the \const{next} pointers linking atoms in the
    same bucket are generally incorrect.  If we have not found the
    atom and the current atom table is too small we must either resize
    the table or some other thread is doing that and we wait for the
    resize to complete.  That is why the resize is locked
    (\algoref{resizeatomtable}, lines 4 and 14).  If the atom table
    changed or the current bucked changed (\algoref{atomlookup},
    lines 27 to 29), our lookup may have failed because the table
    was being resized or a new atom was inserted.  We restart the
    search using the latest table and bucket.

    \item Now, if we did not find the atom, it is not in the table.
    We reserve a new atom by allocating it in the dynamic atom
    array (see \algoref{reserveatom}). Next, if we can CAS the reserved
    atom into the $head$ and the table has not changed
    (\algoref{resizeatomtable}, lines 33 to 36) we added a unique atom
    to the table. We make it valid and return it. If something changed,
    we reset the references to zero, invalidating the atom and restart
    the search. This deals with three scenarios: (1) the table was
    resized while we added the atom, (2) someone else inserted the same
    atom or (3) someone else inserted a different atom in the same
    bucket. The last scenario make us redo the lookup and insert for no
    reason, but this only happens when two threads create two different
    atoms in the same bucket which should be rare.
\end{enumerate}

The above describes lookup of an atom, resizing the table and adding a
new atom to the table while maintaining the unique atom-to-string
mapping.

Two issues still need to be addressed. First, AGC may find the atom is
ready to be collected. This is realised by \algoref{invalidateatom},
where we use CAS to clear the valid bit (lines 4 to 7). This, together
with \algoref{bumpref} which is used to return a found atom from the
table ensures that while doing a lookup of an atom that is being
invalidated by AGC either makes the lookup win, cancelling collection by
AGC or AGC wins and a new atom with the same string is created by the
lookup. Note that if AGC wins the atom changes identity. As the old
identity is not in use, this is harmless. Second, we must reclaim old
data structures: (1) tables that have been resized accessible through
the \const{prev} from $atomTable$ and (2) invalidated atoms that are
linked into $invalidAtoms$ in lines 8 and 9 of \algoref{invalidateatom}.
For this, we keep a pointer at the table and bucket being processed in
the thread's local data. These pointers are updated in
\algoref{atomlookup}, lines 7,~12,~18 and~35. At the end of AGC,
\algoref{reclaimatoms} is called. This collects all bucket pointers in
use by all threads and actually reclaims the atom if none of the buckets
in which the atom must appear in any of the tables is referenced by any
thread. Note that the collected bucket pointers is just a snapshot, but
as none of the current buckets contain the atom, new bucket pointers
will never encounter the atom. Likewise, old tables (\const{prev}) that
are not in use by any thread are reclaimed. This step is trivial and not
included in the pseudo code.

\begin{algorithm}[htb]
\begin{algorithmic}[1]
\footnotesize
\Function{bump_ref}{a, references}
  \Loop
  \If{$\Call{cas}{\&a.references, references, references+1}$}
    \State \Return true
  \Else
    \State $references \gets a.references$
    \If{$\neg \Call{is_valid}{references}$}
      \State \Return false
    \EndIf
  \EndIf
  \EndLoop
\EndFunction
\end{algorithmic}

    \caption{Claim an atom as valid}
    \label{algo:bumpref}
\end{algorithm}

\begin{algorithm}[htb]
\begin{algorithmic}[1]
\footnotesize
\State \textbf{global var} $atomTable$

\State
\Function{resize_atom_table}{}
  \State $\Call{lock}{agc}$
  \If{table too full}
    \State $newtable \gets \Call{alloc_atom_table}{atomTable.buckets*2}$
    \State $newtable.prev \gets atomTable$
    \ForAll{atom $a$ in $atomTable$}
      \If{$\Call{is_valid}{a.references}$}
        \State $\Call{add_to_table}{newtable}$
      \EndIf
    \EndFor
  \EndIf
  \State $atomTable \gets newtable$
  \State $\Call{unlock}{agc}$
\EndFunction

\end{algorithmic}

    \caption{Resize atom table}
    \label{algo:resizeatomtable}
\end{algorithm}

\begin{algorithm}[htb]
\begin{algorithmic}[1]
\footnotesize
\State \textbf{global var} $invalidAtoms$

\State
\Function{invalidate_atom}{a, references}
  \State $newrefs \gets references \& \string~valid$
         \Comment{Clear valid bit}
  \If{$\neg \Call{cas}{\&a.references, references, newrefs}$}
    \State \Return false
  \EndIf
  \State $a.next\_invalid \gets invalidAtoms$
  \State $invalidAtoms \gets a$
  \State \Return true
\EndFunction
\end{algorithmic}

    \caption{Invalidate atom (during AGC collect phase)}
    \label{algo:invalidateatom}
\end{algorithm}

\begin{algorithm}[htb]
\begin{algorithmic}[1]
\footnotesize
\State \textbf{global var} $invalidAtoms$
\State \textbf{global var} $atomTable$

\State
\Function{destroy_atoms}{}
  \State $buckets = \Call{atom_buckets_in_use}{}$\Comment{Collects $LD.atomBucket$ of threads}
   \ForAll{atom $a$ in $invalidAtoms$}
     \If{$\Call{destroy_atom}{a, buckets}$}
       \State remove $a$ from $invalidAtoms$
     \EndIf
   \EndFor
  \State \Call{free}{buckets}
\EndFunction

\State
\Function{destroy_atom}{a, buckets}
  \State $t \gets atomTable$
  \State $key \gets \Call{hash}{a.string}$

  \While{$t != NULL$}
    \State $v = key \& (t.buckets-1)$
    \If{$\&t.table[v]$ \textbf{in} $buckets$}
      \State \Return false\Comment{A thread scans this bucket}
    \EndIf
    \State $t \gets t.prev$
  \EndWhile
  \State $a.name \gets NULL$
  \State $a.references \gets 0$
  \State \Return true
\EndFunction
\end{algorithmic}

    \caption{Reclaim invalidated atoms (during AGC collect phase)}
    \label{algo:reclaimatoms}
\end{algorithm}

\begin{algorithm}[htb]
\begin{algorithmic}[1]
\footnotesize
\State \textbf{global var} $atomArray$

\State
\Function{reserve_atom}{}
  \Loop
    \ForAll{$a$ \textbf{in} $atomArray$}
      \State $refs \gets a.references$
      \If{$\Call{is_free}{refs} \bigwedge
	   \Call{cas}{\&a.references, refs, refs|\const{reserved}}$}
	\State \Return $a$
      \EndIf
    \EndFor
    \State Add new block to $atomArray$ (locked)
  \EndLoop
\EndFunction
\end{algorithmic}

    \caption{Reserve a new atom}
    \label{algo:reserveatom}
\end{algorithm}

\section{Evaluation}
\label{sec:evaluation}

We evaluated the atom table using an artificial test that stresses the
atom table to the limit. Although many applications hardly stress the
atom table, we also identified scenarios from existing applications
where the new atom table significantly improves performance.

For the artificial test we enumerate all answers of the ISO predicate
\predref{sub_atom}{5} where the first argument is instantiated to an
atom consisting of the (Unicode) characters 0..1000. This tests looks up
502,503 atoms. The test is run on multiple threads concurrently. The
hardware is a dual Intel Xeon E5-2650 CPU system ($2\times8 = 16$ cores,
32~threads) running Ubuntu~14.04. We ran the tests in four conditions,
comparing version 6.5.1 (prior to conservative AGC) to 7.3.20 and both
while collecting the volatile atoms and pre-allocating these atoms,
testing only lookup. The results are shown in \tabref{acgtiming}. We
make the following observations:

\begin{itemize}
   \item Concurrent lookup (rows $13 \ldots 24$) shows that, if atom lookup
   is dominant, there is no speedup from using multiple cores when
   using a lock based atom table.  Our lock-free version shows good
   scalability up to 16 threads (the number of physical cores).
   \item With AGC reclaiming the volatile atoms (rows $1 \ldots 12$) we see
   a similar reduction of the total process CPU usage, but a
   much smaller reduction in wall time usage.  The AGC \textit{time}
   column gives a hint.  AGC time is small in the old version, where
   marking is done by the threads in parallel.  It is high in the new
   version, where the AGC thread performs all the marking.
\end{itemize}

\begin{table}
\begin{tabular}{r|r|rr|rrr}
& & \multicolumn{2}{c|}{\bf Time (sec)} & \multicolumn{3}{c}{\bf Atom GC} \\
Row & \# Threads & Process & Wall & \# Invocations & Reclaimed bytes & Time \\
\hline\noalign{\vskip -5pt}
\multicolumn{7}{c}{\bf 6.5.1, AGC active} \\[-7pt]
\hline
1 & 1 & 0.657 & 0.657 & 49 & 660,818,627 & 0.049 \\
2 & 2 & 3.587 & 2.140 & 85 & 1,167,071,837 & 0.159 \\
3 & 4 & 10.725 & 3.965 & 169 & 2,299,293,054 & 0.339 \\
4 & 8 & 33.082 & 5.098 & 183 & 2,304,928,223 & 0.380 \\
5 & 16 & 117.574 & 8.295 & 54 & 719,046,656 & 0.176 \\
6 & 32 & 429.078 & 30.666 & 849 & 9,388,585,427 & 3.121 \\
\hline\noalign{\vskip -5pt}
\multicolumn{7}{c}{\bf 7.3.20, AGC active} \\[-7pt]
\hline
7 & 1 & 0.632 & 0.633 & 49 & 660,941,810 & 0.049 \\
8 & 2 & 1.506 & 0.788 & 98 & 668,102,982 & 0.110 \\
9 & 4 & 3.018 & 0.803 & 49 & 682,083,694 & 0.215 \\
10& 8 & 8.351 & 1.648 & 238 & 2,783,054,946 & 4.987 \\
11& 16 & 20.365 & 4.791 & 491 & 9,103,288,495 & 19.816 \\
12& 32 & 45.590 & 12.369 & 811 & 18,112,260,091 & 44.880 \\
\hline\noalign{\vskip -5pt}
\multicolumn{7}{c}{\bf 6.5.1, atoms pre-allocated} \\[-7pt]
\hline
13&1 & 0.746 & 0.746 & 0 & 0 & 0.000 \\
14&2 & 3.554 & 2.067 & 0 & 0 & 0.000 \\
15&4 & 9.273 & 3.439 & 0 & 0 & 0.000 \\
16&8 & 27.471 & 4.009 & 0 & 0 & 0.000 \\
17&16 & 117.049 & 7.918 & 0 & 0 & 0.000 \\
18&32 & 296.947 & 21.847 & 0 & 0 & 0.000 \\
\hline\noalign{\vskip -5pt}
\multicolumn{7}{c}{\bf 7.3.20, atoms pre-allocated} \\[-7pt]
\hline
19&1 & 0.595 & 0.595 & 0 & 0 & 0.000 \\
20&2 & 1.708 & 0.876 & 0 & 0 & 0.000 \\
21&4 & 2.454 & 0.715 & 0 & 0 & 0.000 \\
22&8 & 4.811 & 0.718 & 0 & 0 & 0.000 \\
23&16 & 10.851 & 0.687 & 0 & 0 & 0.000 \\
24&32 & 28.506 & 1.188 & 0 & 0 & 0.000 \\
\end{tabular}
    \caption{AGC performance for old (6.5.1) and new (7.3.20) versions
	     of	SWI-Prolog.  Note that the last row of each section
	     relies on hyper-threading.}
    \label{tab:acgtiming}
\end{table}

The first real-world evaluation was performed using
ClioPatria\footnote{\url{http://cliopatria.swi-prolog.org}}. ClioPatria
is a linked data platform running on SWI-Prolog. Node identifiers (IRIs)
are represented as atoms. Likewise, RDF \jargon{literals} are
represented as atoms and a $token \rightarrow literal$ index is created
to allow for full text search. The Linked Politics project converted the
European parliament speeches to RDF, creating 26~million triples that
require 9 million atoms to represent as described above. We timed the
loading time. ClioPatria loads the different sources (graphs) in
parallel. The test ran on the same hardware as above, using 32~threads
for loading the data. The results are shown in \tabref{cliopatria}.

\begin{table}
\begin{center}
\begin{tabular}{lrrrrr}
     & \bf Graphs & \bf Triples    & \bf Wall time & \bf CPU \% & \bf CPU time \\
\hline
\bf 7.2.3 (conservative AGC; Linux) & 47 & 26,192,652 & 1667.73   & 1641   &  27365 \\
\bf 7.3.20 (lock-free AGC; Linux)  & 47 & 26,192,652 &  861.85   & 1409   &  12140 \\
\end{tabular}
\end{center}
    \caption{ClioPatria load time for 26M triples.  Times are in seconds.}
    \label{tab:cliopatria}
\end{table}

The second real-world evaluation was performed using the SecuritEase
stock-broking system running on SWI-Prolog. The application was placed
under a representative load, decoding Financial Information eXchange
(FIX) messages. For the test, 500~client requests were processed. The
time to service each request was logged. Each test cycle used the same
FIX message workload and client request workload. The hardware used for
this test is an Intel i7 2720-QM CPU system (4~cores, 8~threads).
Results were obtained for Gentoo Linux 4.1.15 and Windows Server
2008~R2. The results are presented in table \tabref{fix}. Using the
lock-free atom table, the mean time to service requests was noticeably
reduced. Of particular note is the low variance in timings using the
lock-free atom table.

\begin{table}
\begin{tabular}{lrrr}
                              & \bf Mean Time (ms) & \bf Stddev & \bf Max Time \\
\hline
\bf 7.2.3 (conservative AGC; Linux)      &  24.82      &  6.22   &  54.91 \\
\bf 7.3.20 (lock free AGC; Linux)        &  16.66      &  0.42   &  17.99 \\
\hline
\bf 7.2.3 (conservative AGC; Windows)    &  41.04      & 22.10   & 259.71 \\
\bf 7.3.20 (lock free AGC; Windows)      &  24.32      &  0.50   &  26.59 \\
\end{tabular}
    \caption{Time to answer FIX messages}
    \label{tab:fix}
\end{table}

\section{Conclusions}
\label{sec:conclusions}

We have presented a practical and portable approach to implement
lock-free access to the symbol table for concurrent dynamic languages
such as Prolog, Erlang or Ruby. Lookup of existing atoms
scales nearly perfect up to 16~threads on 16~physical cores. Atom
garbage collection only cause thread heap expansion and garbage
collection to suspend. Performance can be further enhanced by using
multiple threads for the marking phase and more fine-grained locks that
protect the marking phase.

\paragraph{Acknowledgements}

We thank Paulo Moura, Joachim Schimpf and Paul Tarau for their input for
\tabref{agcsupport}. The development of SWI-Prolog is supported by the
Dutch national program COMMIT/.

\bibliographystyle{acmtrans}
\bibliography{symboltable}

\label{lastpage}

%

\end{document}